\documentclass{ws-procs9x6-cpt19}
\usepackage{graphicx}
\usepackage{subfigure}

\begin{document}

\newcommand{\refeq}[1]{(\ref{#1})}
\def\etal {{\it et al.}}

\title{CPT-violating Effects on Neutron Gravitational Bound States}

\author{Zhi Xiao}

\address{Department of Mathematics and Physics, North China Electric Power University, Beijing 102206, China}

\begin{abstract}
Analytical solutions with effective CPT-violating spin--gravity corrections
to the neutron's gravitational bound states
are obtained.
The helicity-dependent phase evolution
due to $\vec{\sigma}\cdot\vec{\tilde{b}}$
and $\vec{\sigma}\cdot\hat{\vec{p}}$ couplings
not only leads to spin precession,
but also to transition-frequency shifts
between different gravitational bound states.
Utilizing transition frequencies
measured in the qBounce experiment,\cite{GRS}
we obtain the rough bound  $|\vec{\tilde{b}}|<6.9\times10^{-21}$GeV.
Incorporating known systematic errors
may lead to more robust and tighter constraints.
\end{abstract}

\bodymatter

\section{Introduction}
Lorentz-violating matter--gravity couplings\cite{MGC}
open a broad and interesting avenue for testing Lorentz symmetry.
Recently,
spin-independent Lorentz-violating neutron--gravity couplings
have been thoroughly studied\cite{Escobar1}
in an attempt to analyze the GRANIT experiment.\cite{UCNILL}
However,
to our best knowledge,
an extensive study of spin-dependent fermion--gravity couplings
is still under development.\cite{AKZL}
Here,
we provide a first glimpse
at CPT-violating spin-dependent neutron--gravity couplings.
A more detailed and complete analysis can be found in Ref.\ \refcite{ZXCPTV}.

\section{LV corrections due to spin-dependent interactions}
\label{aba:sec1}

The main vertical hamiltonian
after averaging over the horizontal degrees of freedom
is\cite{Escobar1}
\begin{eqnarray}\label{HLVEff}&&
\hat{H}=\frac{\hat{p}_z^2}{2m_I}+m_{G}\,g\,z-\vec{\sigma}\cdot\vec{\tilde{b}}\,(1+\Phi_0),
\end{eqnarray}
where we have started with the hamiltonian in Ref.\ \refcite{YuriEPI}
and performed a series of redefinitions and approximations.
The stationary solution of \refeq{HLVEff} is
\begin{eqnarray}\label{Spin-HorizT}&&
\Psi_\perp(t,z)=\frac{1}{\sqrt{2}}\left(
              \begin{array}{c}
                \left[\cos(\frac{\Omega}{2}t)+ir_+\sin(\frac{\Omega}{2}t)
                \right]e^{-\frac{i\omega}{2}t} \\
                \left[\cos(\frac{\Omega}{2}t)-ir_-\sin(\frac{\Omega}{2}t)
                \right]e^{+\frac{i\omega}{2}t}\\
              \end{array}
            \right)\phi_n(z)e^{-iE_nt},
\end{eqnarray}
where $\phi_n(z)\equiv\frac{\mathrm{Ai}[\frac{z}{L_c}-x_{n+1}]}{L_c^{1/2}|\mathrm{Ai}'[-x_{n+1}]|}$,
$\Omega\equiv\sqrt{\omega^2+4\omega B_0\cos\theta+4B_0^2}$,
and $r_{\pm}\equiv\left[\omega+2B_0(\cos\theta\pm\sin\theta{e^{-i\phi}})\right]/{\Omega}$;
the initial state is assumed to be an eigenstate of $\sigma_x$,
$|X\uparrow\rangle$.

\begin{figure}
\centering
 \subfigure[Probability to remain in state $|X\uparrow\rangle$]{\includegraphics[width=60mm]{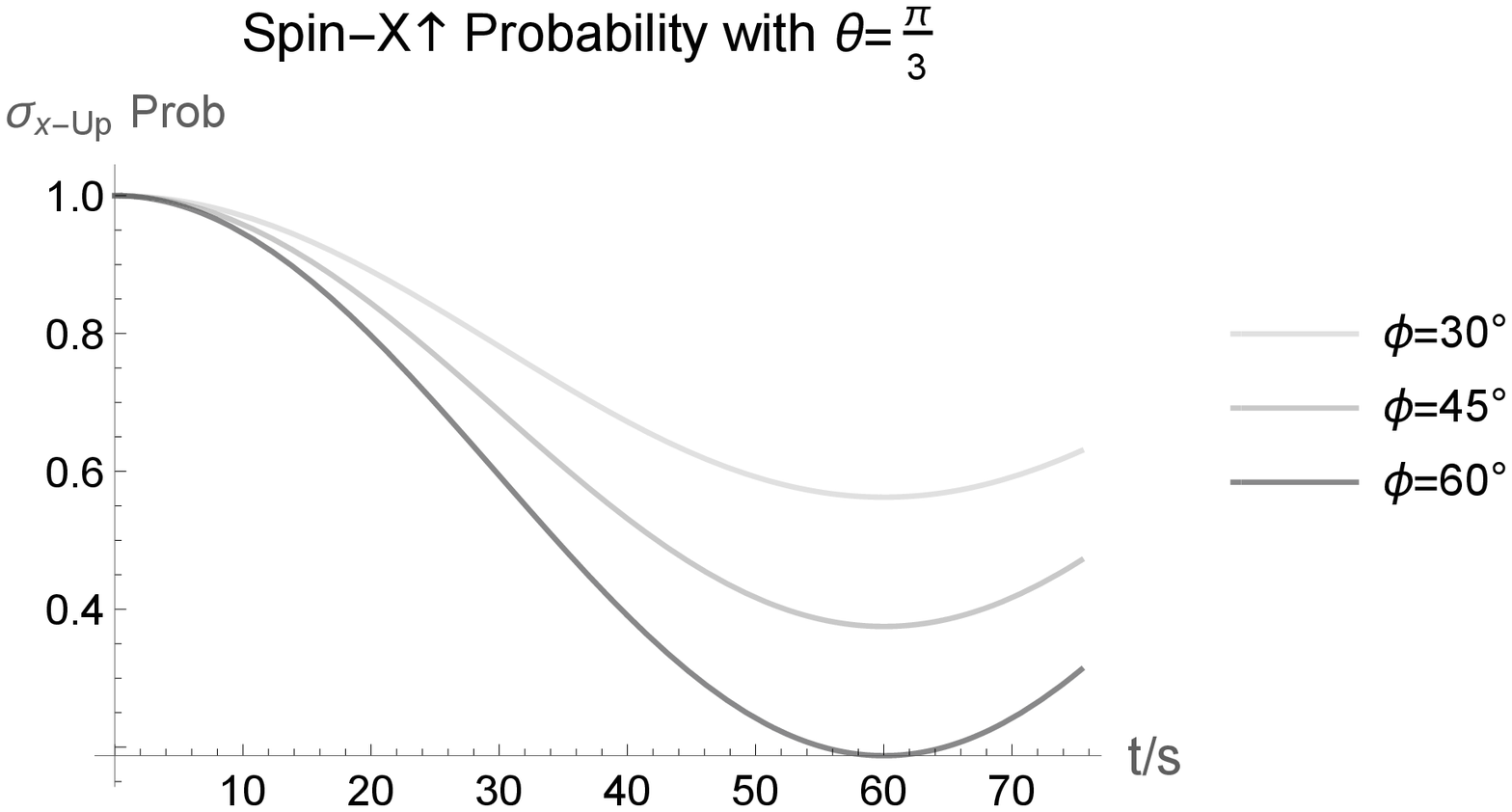}\label{ProbX}}
  \hspace{0.2in}
 \subfigure[Spin-vector rotation]{\includegraphics[width=45mm]{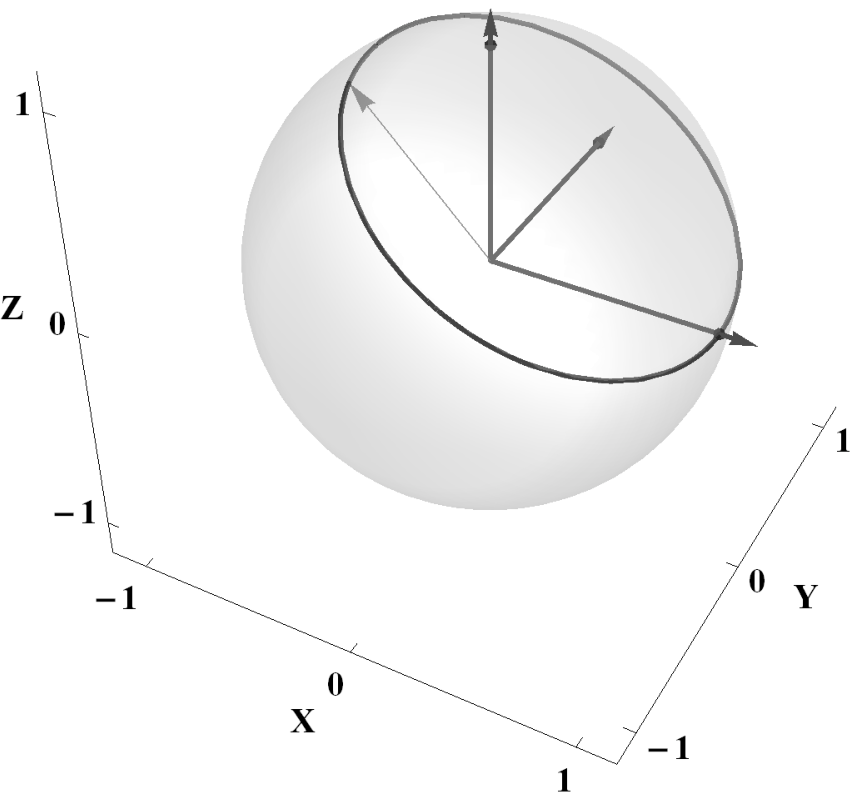}\label{SpinX}}
       \caption{\footnotesize\label{SpinProbX} 
        (a) Probability to remain in the initial state $|X\uparrow\rangle$.
We have chosen an unrealistically large $B_0=\pi/120$
to make the probability variation apparent.
The other parameters are $\theta=\pi/3$ and $\omega=0$;
the $\phi$ dependence is clearly discernible from the three curves.
        (b) Evolution of the spin vector
for the initial state $|X\uparrow\rangle$.
Again, an unrealistically large $B_0=\pi/13$ has been chosen.
The other parameters are $\theta=\pi/6$ and $\phi=\omega=0$.}
\end{figure}

The probability profile for unrealistically large $B_0\equiv|\vec{\tilde{b}}|$
as well as the spin-precession on the Bloch sphere
are shown in Fig.\ \ref{SpinProbX}.
The eigensolution of
$\hat{H}=\hat{p}_z^2/(2m_I)+m_G\,g\,z-\bar{b}_\mathrm{eff}\,(1+\Phi_0)\,\sigma_z\hat{p}_z/m_I$ is
\begin{eqnarray}\label{Spin-Momentum}
\Psi(t,z)=\left(
              \begin{array}{c}
                c_1~e^{i[\bar{b}_\mathrm{eff}(1+\Phi_0)]z} \\
                c_2~e^{-i[\bar{b}_\mathrm{eff}(1+\Phi_0)]z}\\
              \end{array}
            \right)\mathrm{Ai}[\frac{z}{L_c}-x_{n+1}]e^{-i\{E_n-\frac{[\bar{b}_\mathrm{eff}(1+\Phi_0)]^2}{2m_I}\}t},
\end{eqnarray}
where the parity-odd nature of $\vec{\sigma}\cdot\hat{\vec{p}}$
again dictates opposite phase evolutions
for the two helicity components,
and $c_1$ and $c_2$ are constants to be determined.

For $\hat{H}_{LV}=-\vec{\sigma}\cdot\vec{\tilde{b}}\,(1+\Phi_0+gz)-\bar{b}_\mathrm{eff}\,(1+\Phi_0)\,\sigma_z\hat{p}_z/m_I$,
we can use the matrix elements
\begin{eqnarray}\label{MatrixEle}&&
\left(
  \begin{array}{cc}
    \langle{n+}|\hat{H}_{LV}|n,+\rangle & \langle{n+}|\hat{H}_{LV}|n,-\rangle \\
    \langle{n-}|\hat{H}_{LV}|n,+\rangle & \langle{n-}|\hat{H}_{LV}|n,-\rangle \\
  \end{array}
\right)\nonumber\\&&~~~~~~
=-B_0\left[(1+\Phi_0)+\frac{2}{3}gL_cx_{n+1}\right]
\left(
  \begin{array}{cc}
    \cos\theta & \sin\theta e^{-i\phi} \\
    \sin\theta e^{i\phi} & -\cos\theta \\
  \end{array}
\right)
\end{eqnarray}
to calculate the shift in the eigenenergies
\begin{eqnarray}\label{EigenEM}
\delta{E}_n=\mp B_0\left[(1+\Phi_0)+\frac{2}{3}gL_cx_{n+1}\right],
\end{eqnarray}
where the upper or lower sign depends on whether the spin state is
$|\hat{n}+\rangle=(e^{-i\phi}\cos\tfrac{1}{2}\theta,\,\sin\tfrac{1}{2}\theta)$ or
$|\hat{n}-\rangle=(\sin\tfrac{1}{2}\theta,\,-e^{i\phi}\cos\tfrac{1}{2}\theta)$,
respectively.
From (\ref{EigenEM}),
the transition-frequency shift is given by
$\delta\nu_{mn}^\pi
=\mp\frac{2g}{3h}\,|\vec{\tilde{b}}|\,L_c(x_{m+1}-x_{n+1})$.
Comparing $\delta\nu_{mn}^\pi$
with the precisely measured frequencies in the qBounce experiment,\cite{GRS}
we obtain the rough upper bound
$|\vec{\tilde{b}}| < 6.946\times10^{-21}\,$GeV.

\section{Summary}
In this work,
we have discussed CPT-violating spin--gravity corrections
on the neutron's gravitational bound states.
With several analytical solutions,
we have demonstrated
that the phase evolution depends on the helicity of the wave-function components.
The resulting phenomena are spin precession
and $\theta$- and $\phi$-dependent probability variations.
Using degenerate perturbation theory,
we have also calculated the transition-frequency shift.
Comarison with the measurements in Ref.\ \refcite{GRS}
has yielded the rough bound $|\vec{\tilde{b}}|<6.9\times10^{-21}\,$GeV,
which can be improved further
if systematic errors from known physics are taken into account,
or if polarized neutrons are used in the future.

\section*{Acknowledgments}
The author appreciates valuable encouragement
and helpful discussions with M.\ Snow and A.\ Kosteleck\'y
as well as help from many others.
This work is partially supported by the National Science Foundation of China
under grant no.\ 11605056, no.\ 11875127, and no.\ 11575060.

\end{document}